\documentclass[fleqn,twoside]{article}
\usepackage{epsfig}
\usepackage[headings]{espcrc2}

\RequirePackage{xspace}

\readRCS
$Id: sciolla.tex,v 1.2 2005/09/16 14:30:29 sciolla Exp $
\usepackage{graphicx}
\usepackage[figuresright]{rotating}

\def\Bz    {\ensuremath{B^0}}
\def\B     {\ensuremath{B}}
\def\Bbar  {\kern 0.18em\overline{\kern -0.18em B}{}}
\def\Bb    {\ensuremath{\Bbar}}
\def\Bzb   {\ensuremath{\Bbar^0}}
\def\Bz    {\ensuremath{B^0}}
\def\Dm    {\ensuremath{{\rm \Delta}m}} 
\def\Dt    {\ensuremath{{\rm \Delta}t}}

\newcommand{\Btag}{B_{\mathrm{tag}}}

\newcommand{\jprBase}        {Phys.\ Rev.\xspace}
\newcommand{\jplBase}        {Phys.\ Lett.\xspace}
\newcommand{\npBase}         {Nucl.\ Phys.\xspace}
\newcommand{\plb}       [1]  {\jplBase\ B~{\bf #1}}
\newcommand{\jprd}      [1]  {\jprBase\ D~{\bf #1}}
\newcommand{\npb}       [1]  {\npBase\ B~{\bf #1}}

\newcommand{\AmS}{{\protect\the\textfont2
  A\kern-.1667em\lower.5ex\hbox{M}\kern-.125emS}}

\title{
Recent measurements of sin2$\beta$ at BaBar} 

\author{G. Sciolla \address[MIT]{Massachusetts Institute of Technology, 
    Department of Physics,   \\
        Room 26-443,  77 Massachusetts Ave., Cambridge MA 02139 }%
        \thanks{Representing the BaBar Collaboration.}
}       

\runtitle{Recent measurements of sin2$\beta$ at BaBar} 
\runauthor{G. Sciolla}

\begin{document}

\begin{abstract}

The angle $\beta$ is the most accurately measured 
quantity that determines the Unitarity Triangle. In this article I review 
the various measurements of this angle performed by the 
BaBar Collaboration, and discuss their implications in the search 
for new physics. 

\vspace{1pc}
\end{abstract}

\maketitle

\section{INTRODUCTION}

The BaBar detector\cite{babar} is located at the $e^+e^-$ asymmetric B-factory at SLAC. 
The main physics goal of BaBar is to study CP-violation in the $B$ system and  
provide a quantitative test of the $CP$ sector of the Standard Model. 

The determination of the angle $\beta$ through the study of  \Bz\ decays 
plays a key role in this test. 
The measurement of $\beta$ from decays mediated by $b\to c\overline{c}s$ 
tree level diagrams allows for a precise
test of CP violation in the Standard Model and provides the most precise constraint in the 
determination of the parameters $\rho$ and $\eta$. In addition,  
the measurement of the same angle in final states mediated 
by penguin decays and $b\to c\overline{c}d$ diagrams can be used to look for new physics.

All measurements presented in this article are based on a dataset of $232 \times 10^6$ \B\Bb\  events. 


\section{ THE EXPERIMENTAL TECHNIQUE }

At the $B$-factories, $CP$ violation is studied through the measurement of the time dependent $CP$ 
asymmetry, $A_{CP}(t)$. This quantity is defined as 
\begin{equation}
A_{CP}(t) \equiv \frac{N(\Bzb(t)\to f_{CP}) - N(\Bz(t)\to f_{CP})} {N(\Bzb(t)\to f_{CP}) + N(\Bz(t)\to f_{CP})},  
\label{acpt}
\end{equation}
where $N(\Bzb(t)\to f_{CP})$ is the number of \Bzb\ that decay into the $CP$-eigenstate $f_{CP}$ after a time $t$. 

In general, this asymmetry can be expressed as the sum of two components: 
\begin{equation}
A_{CP}(t) =  S_f \sin(\Delta m t) - C_f \cos(\Delta m t), 
\label{acpt2}
\end{equation}
where $\Delta m$ is the difference in mass between $B$ mass eigenstates. 

When only one diagram contributes to the final state, the cosine term vanishes. 
For decays such as $B\to\ J/\psi K^0$, $S_f = -\eta_f \times sin2\beta$, 
%
where  $\eta_f$ is the  $CP$ eigenvalue of the final state, negative 
for charmonium + $K_S$ and positive for  charmonium + $K_L$. 
%
It follows that 
\begin{equation}
A_{CP}(t) = -\eta_f \sin2\beta \sin(\Dm \Dt),  
\label{acpt5}
\end{equation}
which shows how the angle $\beta$ is directly and simply measured by the amplitude 
of the time dependent $CP$ asymmetry.

The measurement of $A_{CP}(t)$ utilizes  decays of the $\Upsilon (4S)$ into two neutral $B$ mesons, 
of which one ($B_{CP}$) can be completely  reconstructed  into a $CP$ eigenstate, 
while the decay products of the other ($\Btag$) identify its flavor at decay time. 

%
%
%
 
The time $t$ between the two $B$ decays is determined by reconstructing the two $B$ decay vertexes. 
The $CP$ asymmetry amplitudes are determined from an unbinned maximum likelihood fit 
to the time distributions separately for events tagged as \Bz\ and \Bzb .

\section{INDEPENDENT MEASUREMENTS OF THE ANGLE $\beta$}

\begin{figure*}[th]
\begin{center}
\psfig{file=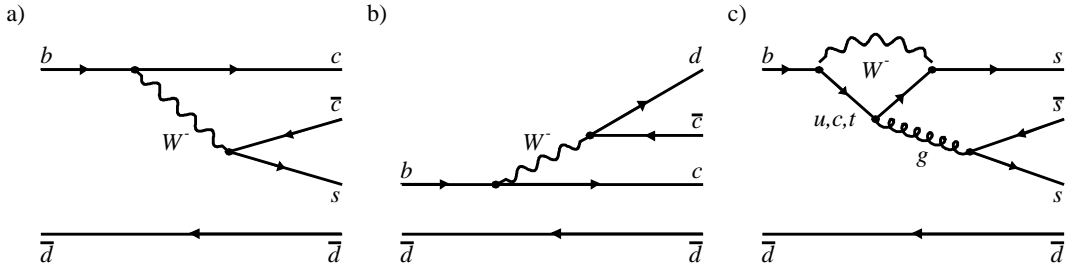,width=0.9\linewidth}
\end{center}
\vspace{-0.4cm}
\caption{Feynman diagrams that mediate the \Bz\ decays used to measure the angle $\beta$: 
a) $\Bz\to\mbox{charmonium}+K^0$; b) $\Bz\to  D^{(*)+}D^{(*)-}$; c) penguin dominated $B$ decays.}  
\label{fey}
\end{figure*}

The angle $\beta$ can be independently measured through the three types of \Bz\ decays illustrated in fig. 
\ref{fey} and discussed in the following. 

\subsection{$\Bz\to\mbox{charmonium}+K^0$}	
These decays,
known as ``golden modes'', 
are dominated by a tree level diagram $b\to c\overline{c}s$ with 
internal $W$ boson emission (fig. \ref{fey}-a). 
The leading penguin diagram contribution to the final state has the same weak phase as the tree diagram, 
and the largest term with different weak phase is a penguin diagram contribution suppressed by $O(\lambda^2)$. 
This makes $C_f=0$ in equation~\ref{acpt2} a very good approximation. 

Besides the theoretical simplicity, these modes also offer experimental advantages because of  their  
relatively large branching fractions ($\sim 10^{-4}$) 
and the presence of the narrow $J/\psi$ resonance in the final state,
which provides a powerful rejection of combinatorial background.

The $CP$ eigenstates considered for this analysis are $J/\psi K_S$, $\psi$(2S)$K_S$, 
$\chi_{c1}K_S$, $\eta_cK_S$ and $J/\psi K_L$. 

\begin{figure}[th]
\psfig{file=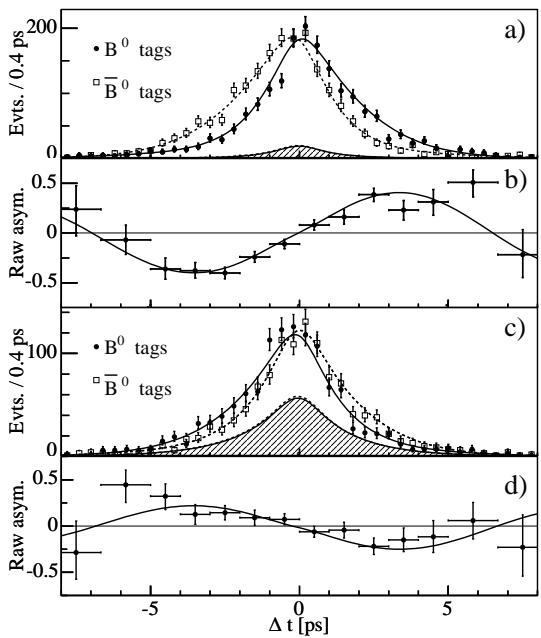,width=0.45\textwidth}
\vspace{-0.6cm}
\caption{Measurement of $\sin 2\beta$ in the ``golden modes'' by BaBar.
Figure a) shows the time distributions for events tagged as \Bz\ (full dots) or \Bzb\ (open squares) 
in $CP$ odd  (charmonium $K_S$) final states. Figure b) shows the  
corresponding raw $CP$ asymmetry with the projection of the unbinned maximum likelihood fit superimposed. 
Figure c) and d) show  the corresponding distributions for $CP$ even ($J/\psi K_L$) final states. }
\vspace{-0.4cm}
\label{babarsin2b-2}
\end{figure}

The asymmetry between the two \Dt\ distributions, clearly visible in figure \ref{babarsin2b-2},  
is a striking manifestation of $CP$ violation in the $B$ system. 
The same figures also display the corresponding raw $CP$ asymmetry with the 
projection of the unbinned maximum likelihood fit superimposed. 
The results of the fit is $\sin 2\beta=0.722\pm 0.040\pm 0.023$\cite{sin2b}. 
The main sources of systematic errors are uncertainties in the background level and characteristics, 
in the parameterization of the time resolution,  
and in the measurement of the mis-tag fractions. Most of these uncertainties 
will decrease with additional statistics. 

\begin{figure}[htb]
   \psfig{file=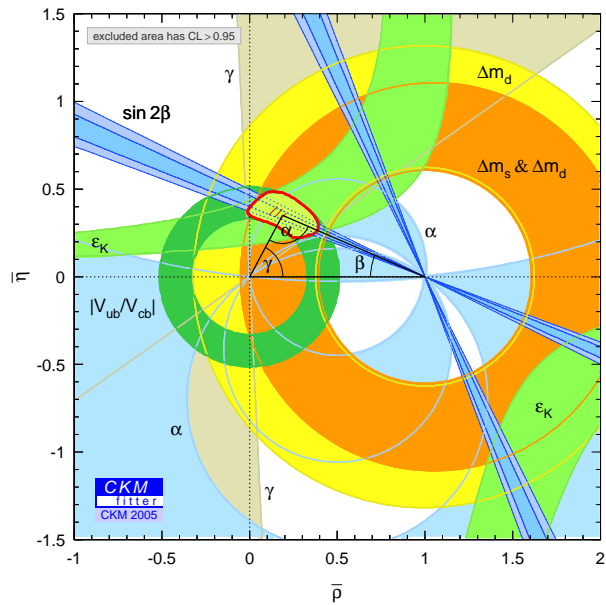,width=0.5\textwidth}
\vspace{-1.4cm}  
 \caption{Comparison between the direct measurement of $\sin 2\beta$ and the constraints on the 
    apex of the Unitarity Triangle from the measurement of the sides and $CP$ violation in the kaon system. }
   \label{CKMFitter1}
\vspace{-0.7cm}
\end{figure}

The world average value for  $\sin 2\beta$, heavily dominated by the results from BaBar\cite{sin2b} and 
Belle\cite{sin2bbelle}, 
is $\sin 2\beta = 0.725\pm 0.037$. 
This value can be compared with the indirect constraints on the apex of the Unitarity Triangle  originating from 
measurements of $\epsilon_K$, $|V_{ub}|$, $|V_{cb}|$, \Bz\ and $B_S$ mixing
as described, for example, in reference~\cite{CKMFitter}. 
The comparison, illustrated in figure \ref{CKMFitter1}, shows excellent agreement between the 
measurements, indicating  that the observed $CP$ asymmetry in $\Bz\to\mbox{charmonium}+K^0$ 
is consistent with the predictions of the CKM mechanism.

The measurement of $sin 2 \beta$ leads to a four-fold ambiguity in the determination of the angle $\beta$. This 
ambiguity can be  reduced to a two-fold ambiguity measuring the sign of $cos 2 \beta$, which can 
be measured  in a time-dependent angular analysis of 
the decays $J/\psi K^{*0}$($K^{*0}\to K_S \pi^0$).
An analysis~\cite{cos2b} based on 81.9 fb$^{-1}$ measures 
$cos 2 \beta = +2.72 ^{+0.50}_{-0.79}\pm 0.27$, which  determines the sign of $cos 2 \beta $ to be positive 
at the 86.6\% C.L., in agreement with the Standard Model expectation. 

\subsection{$\Bz\to D^{(*)+}D^{(*)-}$}
The decays $\Bz\to D^{(*)+}D^{(*)-}$ are 
dominated by a tree level diagram $b\to c\overline{c}d$ with
external $W$ boson emission (fig. \ref{fey}-b).
In the Standard Model, the leading penguin diagram contribution to this final state is expected to be small. 
The penguin-induced correction has been estimated using factorization and heavy quark symmetry to be 
about 2\%~\cite{pengD}. 
New physics could enhance the penguin contribution and would lead to a measurement of time dependent $CP$ asymmetry 
substantially different from that measured in $\Bz\to\mbox{charmonium}+K^0$ decays.  

BaBar measures $A_{CP}(t)$ in three different channels: $D^{*+}D^{*-}$\cite{d*d*}, 
$D^{+}D^{*-}$ and $D^+D^-$\cite{d*d}. 
Since 
two different diagrams can contribute to the final state 
in these decays,  
both the sine ($S_f$) and cosine ($C_f$) terms have to be 
extracted in the fit for  $A_{CP}(t)$. 
The extraction of the coefficients S and C is straightforward for the channel $D^{+}D^{-}$, 
which is a  pure CP eigenstate. 
Instead, the decay $\Bz\to D^{*+}D^{*-}$ is an admixture of  CP-odd and CP-even components. 
The CP-odd fraction is measured 
by means of a transversity analysis\cite{d*d*} 
to be $0.125 \pm 0.044 \pm 0.070$. 

The results of the CP fit for $B$ decays to open charm reported in table \ref{tabDD}
are in agreement with the Standard Model expectations.

\begin{table}[htb]
\caption{Measurements of the sine (S) and cosine (C) coefficient in the 
fit of $A_{CP}(t)$ in $b\to c\overline{c}d$ decays.}
\label{tabDD}
\newcommand{\m}{\hphantom{$-$}}
\newcommand{\cc}[1]{\multicolumn{1}{c}{#1}}
\renewcommand{\tabcolsep}{0.9pc} 
\renewcommand{\arraystretch}{1.2} 
\begin{tabular}{@{}lll}
\hline
Channel           & \cc{$\eta_{CP} \times S$} & \cc{$C_f$} \\
\hline
$D^{*+}D^{*-}$    & \cc{$ 0.75 \pm 0.25  \pm 0.03$} & \cc{$0.06 \pm 0.17 \pm 0.03 $} \\
\hline
$D^+D^-$          & \cc{$ 0.29 \pm 0.63   \pm 0.06 $}  & \cc{$ 0.11  \pm 0.35  \pm 0.06 $} \\
\hline
$D^{*+}D^-$       & \cc{$ 0.54   \pm 0.35  \pm 0.07 $} & \cc{$0.09   \pm 0.25  \pm 0.06 $} \\
\hline
$D^{*-}D^+$       & \cc{$ 0.29   \pm  0.33  \pm 0.07 $} & \cc{$0.17   \pm 0.24  \pm 0.04 $} \\
\hline
\end{tabular}\\[2pt]
\vspace{-0.3cm}
\end{table}


\subsection{Penguin dominated $\Bz$ decays }
In the Standard Model, 
final states dominated by $b\to s \overline{s} s $ or $b\to s \overline{d} d $ decays 
offer a clean and independent way of measuring $sin2\beta$\cite{sPenguin}. 
Examples of these final states are 
$\phi K^0$,  $\eta 'K^0$, $f_0K^0$, $\pi^0 K^0$, $\omega K^0$, $K^+K^-K_S$ and  $K_S K_S K_S$.
These decays are mediated by the gluonic penguin diagram illustrated in figure \ref{fey}-c. 
In presence of physics beyond the Standard Model, new particles such as  
squarks and gluinos, could participate in the loop and affect the time 
dependent asymmetries\cite{phases}.  

The decays $\Bz\to \phi K_S$ are ideal for these studies. In the Standard Model, 
these decays are almost pure $b\to s \overline{s} s $ penguin decays, and their CP asymmetry is 
expected to coincide with the one measured in charmonium + K$^0$ decays within a few percent~\cite{phases}.
Experimentally, this channel is also very clean,  
thanks to the powerful background suppression due to the narrow $\phi$ resonance. Unfortunately, 
the branching fraction for this mode is quite small ($\approx 8\times 10^{-6}$), therefore the 
measurement is affected by a large statistical error. 

The decays $\Bz\to \eta^{\prime}K_S$ are favored by a
larger branching fraction ($\approx 6\times 10^{-5}$).  
In the Standard Model, these decays are also dominated by penguin diagrams; other contributions 
are expected to be small~\cite{etaprimetheory}.  

\begin{figure}[th]
\begin{center}
\psfig{file=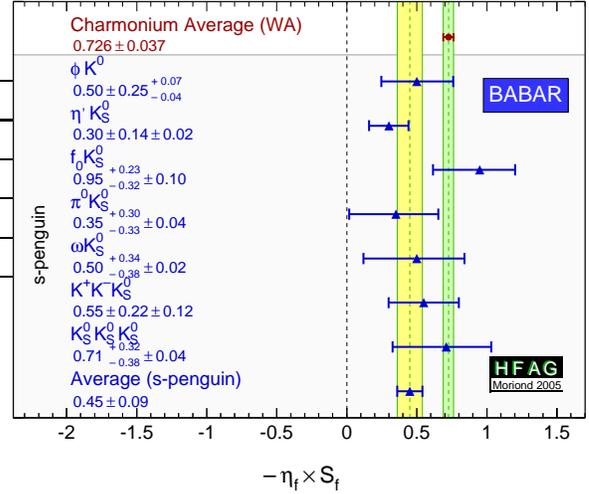,width=0.5\textwidth}
\end{center}
\vspace{-1.4cm}
\caption{BaBar measurements of ``$\sin 2\beta$'' in the penguin dominated  channels. 
The green band indicates the world average of the charmonium + $K^0$ final states $\pm 1 \sigma$;
the yellow band is the average of the s-penguin modes $\pm 1 \sigma$. 
}
\vspace{-0.3cm}
\label{FigurePenguin}
\end{figure}

A summary of the measurements of $A_{CP}(t)$ in penguin modes~\cite{phiks,etaprimeKs,kspi0,f0k0,omegaKs}
by the BaBar experiment is reported in figure \ref{FigurePenguin}. 
The average of  all the penguin modes~\cite{hfag}, shown in yellow in figure \ref{FigurePenguin}, 
is 2.8$\sigma$ away from the value of  $\sin 2\beta$ measured in the golden mode. 
This discrepancy, however, has to be interpreted with caution since each 
 mode is theoretically affected by new physics in different ways.

\section{ SUMMARY AND CONCLUSION  }

The measurement of time-dependent CP asymmetry in \Bz\ decays have provided a crucial test of 
$CP$ violation in the Standard Model. 
The parameter $\sin 2\beta$ is now measured in $b\to c\overline{c}s$ decays 
by BaBar with a precision of 5\%. 
Measurements of time-dependent $CP$ violation asymmetries  in $b\to c\overline{c}d$ 
and in penguin-dominated modes are sensitive to contributions from physics beyond the 
Standard Model. These measurements are still heavily dominated by statistical errors and will 
benefit greatly from additional data. BaBar is planning to double its dataset by 
2006 and quadruple it by 2008. 

\end{document}